\documentclass{aastex62}
\usepackage{booktabs,multirow}
\usepackage{xcolor}
\usepackage{ulem}

\newcommand{\fic}{f_{\mathrm{ic}}}
\newcommand{\fvol}{f_{\mathrm{vol}}}
\newcommand{\rhocl}{\rho_{\mathrm{cl}}}
\newcommand{\rhoic}{\rho_{\mathrm{ic}}}

\graphicspath{{./}{figures/}}

\received{2018 November 30}
\revised{2019 January 22}
\accepted{2019 January 30}
\submitjournal{ApJ}

\shorttitle{Inner Photons in AGB Chemistry}
\shortauthors{Van de Sande and Millar}

\begin{document}

\title{The Role of Internal Photons on the Chemistry of the Circumstellar Envelopes of AGB Stars}

\correspondingauthor{T . J. Millar}
\email{tom.millar@qub.ac.uk} 

\author[0000-0002-0786-7307]{M. Van de Sande}
\affil{Department of Physics and Astronomy, Institute of Astronomy, KU Leuven, \\
Celestijnenlaan 200D, 3001 Leuven, Belgium}
\email{marie.vandesande@kuleuven.be}

\author[0000-0001-5178-3656]{T. J. Millar}
\affiliation{Astrophysics Research Centre, School of Mathematics and Physics, \\
Queen's University Belfast, University Road, Belfast BT7 1NN, UK}
\affiliation{Leiden Observatory, Leiden University, PO Box 9513, 2300 RA Leiden, The Netherlands}



\begin{abstract}

Recent high spatial resolution observations of gas and dust in the circumstellar envelopes 
(CSEs) of AGB stars indicate morphologies much more complex than the smooth density 
distributions generated by spherically symmetric, constant mass loss rates. 
In particular, the observation of spiral arcs and disks indicate the likely presence of a 
binary companion which in some cases give rise to the UV photons detected by {\em GALEX}.  
In this {\em Article}, we extend our recent model of the chemistry in a clumpy, porous 
CSE around an AGB star to include the influence of stellar blackbody photons on the CSE chemistry.  
Our results indicate that internal photons, in a clumpy, porous CSE, can alter 
chemistry within a few stellar radii and, for some molecules, alter abundances out to 
several hundred stellar radii. They further suggest that harder radiation from 
companion stars or accretion disks will have a substantial impact on chemistry in the 
dust formation zones and inner CSEs of AGB stars.

\end{abstract}



\section{Introduction} \label{sec:intro}

Astrochemical studies of the the circumstellar envelopes (CSEs) of Asymptotic Giant Branch (AGB) stars 
are of central importance in understanding macroscopic processes that are important in astronomy.
These include the origin of interstellar dust, nucleosynthesis of the elements, the recycling of matter
through mass loss, and the end points of stellar evolution.  The CSEs are, in general, rich in
molecular material, with close to 100 different molecules detected therein, as well as
being efficient factories for the formation of the dust particles that populate the interstellar media of galaxies.
They have long been known to contain large-scale density structures such as arcs and rings 
in both optical/near-IR \citep{mau00,leao06}, and millimeter/sub-millimeter interferometric 
observations \citep{gue99,tru08,agu17}.  More recently, very high spatial resolution 
observations of both line and continuum emission have indicated the presence of structures such as spirals
\citep{mau06,dec15,qui16,hom18}, disks \citep{ker14}, and clumps, including small-scale structures close to 
the stellar photosphere \citep{kho16,wit17,dec18a}. 

The origin of such structures is still debated but 
there is evidence that binary companions may play an important role here \citep{dec15,qui17,ram17}
as well as in shaping the non-spherical morphologies of proto-planetary and planetary nebulae \citep{dem09}.
Furthermore, recent {\em GALEX} detections of UV radiation from AGB stars are argued to arise from 
sources that are either intrinsic, eg. from chromospheres or pulsational shock 
waves \citep{mon17} or extrinsic, e.g. from binary companions or accretion 
disks \citep{sah08, sah11, sah18}.  \citet{ort19} conclude that far-UV emission may be 
extrinsic but that further observations are required to draw a definite conclusion. The detection of 
CI in the O-rich AGB star omi Ceti has also been argued as due to an internal source of 
UV photons \citep{sab18}.

In addition, the presence of `unexpected' molecules,
such as hot H$_2$O in C-rich CSEs \citep{dec10,neu11} and CS, CN and HCN in O-rich 
envelopes \citep{lin88,buj94}, 
as well as the detection of species such as CH$_3$CN \citep{agu15} and NaCN \citep{qui17} 
on angular scales of 1.5--3 arcsec, present a challenge for the traditional chemical kinetic models 
that incorporate a constant velocity and a constant mass-loss rate for the CSE structure.

\citet{cor09} considered chemical models that included enhanced periods of mass-loss to
reproduce the observed, ring-like, molecular distributions in IRC+10216 with some success in that 
their additional shielding and increased shell density tended to restrict chemical evolution to these
higher density rings, but the model is unable to account for the presence of hot water deep in the CSE \citep{dec10}.
Non-thermal equilibrium processes driven by pulsation-induced shocks may provide sufficient H$_2$O \citep{che11,che12}.
Another promising approach was introduced by \citet{agu10} who, on the basis of the 
clumpy structures seen in millimeter emission lines, presented a model that allowed a fraction of
interstellar UV photons to penetrate deep into the CSE free from extinction due to
circumstellar dust grains. This model was able to generate abundant, hot H$_2$O, in the inner CSE
as well as perturb the chemistry in other ways.  More recently, \citet{vds18a}  
presented a more sophisticated porosity formalism that takes into account both an 
enhanced penetration of interstellar UV radiation, and  the relative overdensity of clumps 
in the outflow.
We describe its implementation in our model in Section~\ref{sec:porosity}.

Since clumped gas and dust appears to be present even at scales of the dust condensation radius
in AGB stars \citep{agu15,qui17,dec16,kam17,kam16,dec17,dec18a}, it is natural to consider 
whether the presence of an internal source of UV photons 
can drive a chemistry in the inner regions of the CSE, in regions opaque to interstellar UV.
Furthermore, although AGB stars are cool and are not generally considered as sources of UV 
radiation that can affect CSE chemistry, we note that UV photons, 
such as those detected by {\em GALEX}, can penetrate the innermost CSE, particularly if it is clumpy.  
In this {\em Article}, we consider the simple case of UV photons generated by a 
blackbody at an effective temperature and use our porosity formalism to investigate
the influence of UV emission from a cool AGB star on circumstellar chemistry, deferring a 
discussion of UV photons from a hotter companion to future work.  This work is particularly 
timely in anticipation of the Cycle 6 ALMA Large Programme, ATOMIUM (2018.1.000659.L, P.I. Decin) 
which will use Band 6 observations to study molecules in the dust formation zones of over 
20 late-type stars.

\section{Internal UV and the Porosity Formalism} \label{sec:porosity}

The porosity formalism assumes that the CSE is composed of a stochastic ensemble of small-scale 
density enhancements or clumps, which take up a fraction $\fvol$ of the total volume of the outflow, 
embedded in a rarefied inter-clump component, with a mean density relative to that of the smooth flow 
given by $\fic$. 
The density in the smooth flow is related to $\fvol$ and $\fic$ by:

\begin{equation}
\rho = (1-\fvol) \rhoic + \fvol \rhocl
\end{equation}
where we have suppressed the radial dependencies of the densities, and $\rhocl$ and $\rhoic$ 
are the clump and inter-clump densities, respectively. 
The ratio between the clump size $l(r)$ and $\fvol$ represents the mean-free-path between the 
clumps, $h(r)$, a measure of the porosity of the outflow.
Note that mass conservation within the clumps implies that both the characteristic clump size $l(r)$ and $h(r)$ 
are proportional to $r^{2/3}$.
The specific clumpiness of the outflow is hence fully described by the parameters $\fvol$, 
$\fic$ and $l_*$, the characteristic clump size at the stellar surface.
The larger penetration of interstellar UV photons is taken into account by changing the optical 
depth of the outflow to an effective, clumpy, optical depth \citep{vds18a}.
Our model describes the chemical kinetic evolution of both the clump and inter-clump gas  
and calculates the fractional abundance of species X relative to H$_2$ as a function of radius via:

\begin{equation}
y_{\mathrm{X}} = y_{\mathrm{cl,X}} + \fic (1 - \fvol)(y_{\mathrm{ic,X}}-y_{\mathrm{cl,X}})
\end{equation}

As a first approximation we have used blackbody radiation to determine the internal UV fluxes. 
We have considered whether stellar atmosphere models can provide a more accurate estimate but 
find that the Kurucz\footnote{\url{http://kurucz.harvard.edu/grids.html}}, 
MARCS\footnote{\url{http://marcs.astro.uu.se/}} and 
PHOENIX\footnote{\url{http://osubdd.ens-lyon.fr/phoenix/}} models have either temperatures 
or surface gravities that are too large than those appropriate for well observed 
low-to-intermediate mass AGB stars such as CW Leo and R Dor \citep{cas03,gus08,hus13}. 
Calculated values of log g are -1.4 for IRC+10216 and -1.17 for R Dor, both smaller than 
those offered by stellar atmosphere models.

\begin{figure}[ht!]
\centering
\centering
\includegraphics[scale=0.5]{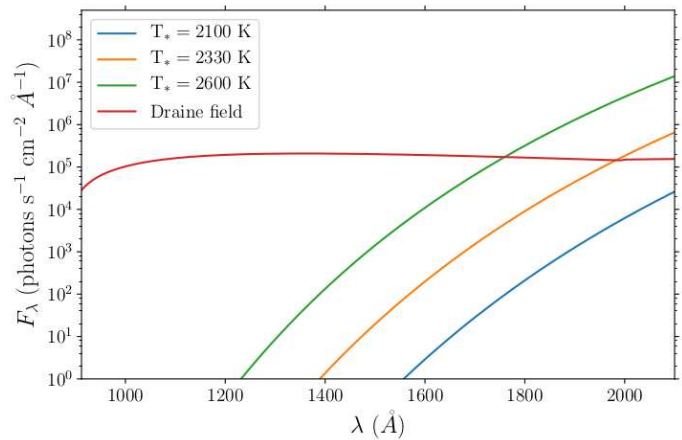}
\caption{The blackbody flux at a distance of 50\,R$_*$ for a stellar radius of 5 $\times$ 
10$^{13}$ cm for temperatures of 2100\,K, 2330\,K, equivalent to IRC+10216, and 2600\,K.
\label{fig:BB}}
\end{figure}

Figure \ref{fig:BB} shows the blackbody photon flux at 2.5 $\times$ 10$^{15}$ cm (50\,R$_*$ 
for IRC+10216) for the wavelength range 912--2100\,\AA, approximately 6--13.6\,eV. Although there are 
negligible photons at the shortest wavelengths, the figure shows that the flux increases rapidly 
at longer wavelengths. We have calculated the photon flux, $F_{*,\mathrm{sc}}(\lambda)$, at a fiducial
radius, $R_{\mathrm{sc}}$ = 50 R$_*$, and used this to derive the unshielded photo-rate coefficients 
with cross-sections
taken from the Leiden Observatory 
Database\footnote{\url{http://www.strw.leidenuniv.nl/~ewine/photo}} \citep{hea17}
as:

\begin{equation}
\alpha^{\mathrm{IP}}_{0,\mathrm{sc}}  =  \int \! F_{*,\mathrm{sc}}(\lambda) \sigma(\lambda) \mathrm{d}\lambda \label{eqn:exact}
\end{equation}

This dataset covers over 120 photoreaction channels in our network. To these, we have used 
a mixture of laboratory measurements and theoretical calculations on electron affinities (EA) to 
estimate the photodetachment rates of around 25 anions \citep{arn91,bla01,bes10,yen10,kum13,sen13}
with cross-sections determined from the formula given by \citet{mil07}:

\begin{equation}
\sigma = \sigma_{\infty} (1 - EA/E)^{0.5}
\end{equation}
where $\sigma_{\infty}$ is the asymptotic cross-section at high energy and E is the photon energy. 
We note that anions do not influence the chemistry of the inner CSE significantly as they undergo 
rapid photodetachement due to their low electron affinities.

For species for which cross-sections are not available, we have scaled the interstellar unshielded rate
by the ratio of the integrated fluxes of stellar to interstellar photons (integrated over 6--13.6\,eV),
a common approach in astrochemical modelling:

\begin{equation} 
\alpha^{\mathrm{IP}}_0  =  \left(\frac{G_*}{G_{\mathrm{IS}}}\right) \alpha^{\mathrm{IS}}_0 \label{eqn:integ}
\end{equation}
where 
\begin{equation}
G = \int_{2068\AA}^{912\AA} \! F(\lambda) \mathrm{d}\lambda
\end{equation}

We note, however, that this approach can 
significantly overestimate the photoionization rates for molecules since their ionization potentials
generally fall at higher energies, where the flux of stellar UV photons is negligible, 
than their bond dissociation energies. For these species, we have therefore
used a scaling factor, determined from comparing photoionization rates calculated from (exact) 
atomic cross-sections (Equation~\ref{eqn:exact}) with those from the integrated approach (Equation~\ref{eqn:integ}), to 
reduce their calculated rates to insignificant values.  In total we include 436 photochannels 
due to internal photons. 

At every radial distance $r$, we calculate the effective dust extinction
in the porosity formalism, A$_{\mathrm{V}}^{\mathrm{eff}}$, in terms of the optical depth for a uniform
outflow, A$_{\mathrm{V}}$.  The equations for both a one-component model, where the inter-clump 
medium is void,
and a two-component model, where mass is distributed in both a clump and inter-clump medium, 
are given in Appendix C of \citet{vds18a}. The internal photon flux is diluted geometrically 
and extinguished by dust. For the dust extinction experienced by internal photons, the dust 
condensation radius is set at 1.9\,R$_*$, or at a temperature close to 1500\,K. 

\begin{figure}[ht!]
\centering
\centering
\includegraphics[scale=0.5]{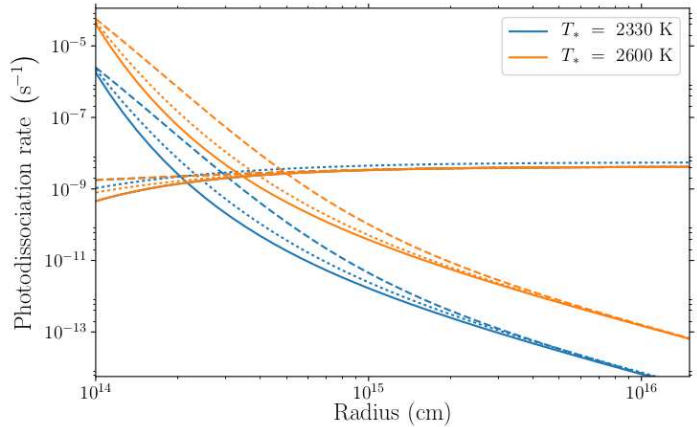}
\caption{The internal and external UV photodissociation rates of SO are plotted as a function of 
radius for a mass-loss rate of 10$^{-7}$ M$_{\odot}$ yr$^{-1}$ for effective stellar temperatures of 2330\,K
(blue), equivalent to IRC+10216, and 2600\,K (orange). Solid curves represent a smooth outflow, dashed curves a
one-component outflow with $\fvol$ = 0.1 and $l_* = 10^{13}$ cm, and dotted curves a two-component outflow 
with $\fvol$ = 0.1, $\fic$ = 0.5 and $l_* = 10^{13}$ cm.
 \label{fig:so_pd}}
\end{figure}

\begin{figure*}[ht!]
\centering
\centering
\includegraphics[scale=0.4]{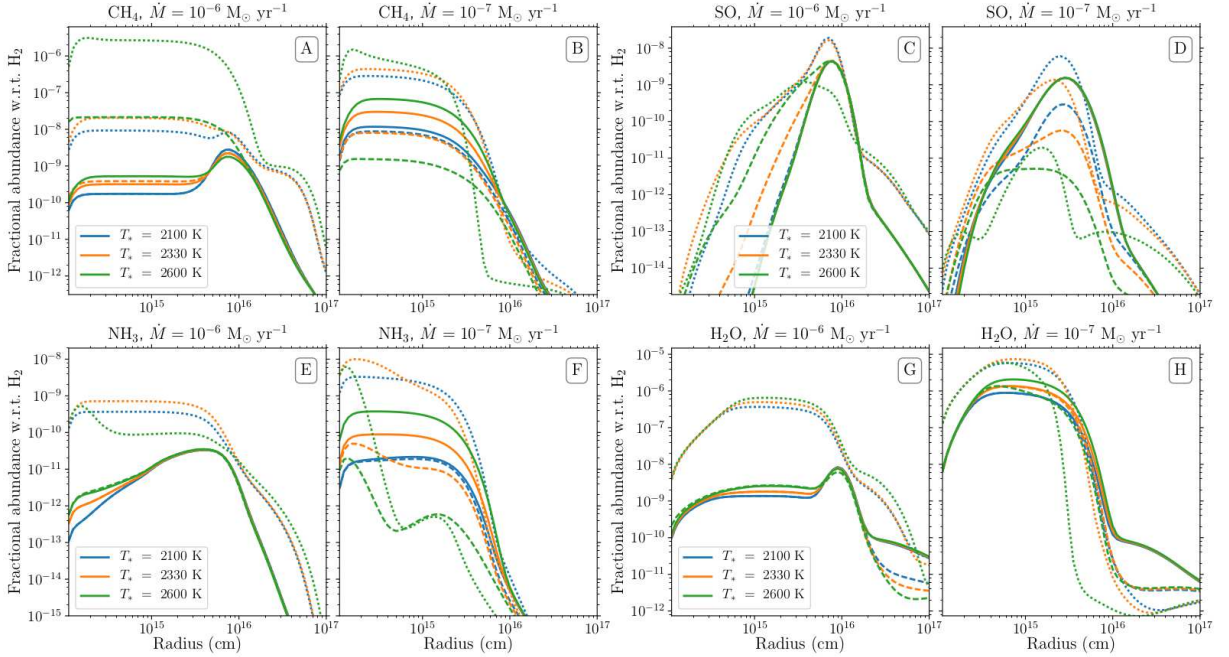}
\caption{Radial fractional abundance of CH$_4$, SO, NH$_3$ and H$_2$O for 2100\,K (blue), 
2330 K (orange) and 2600 K (green) in the C-rich case. 
Solid curves refer to a smooth outflow without internal photons, 
dashed curves to a smooth outflow with internal photons, and dotted curves to a clumpy, 
one-component outflow with internal photons. \label{fig:ch4-so}}
\end{figure*}

As an example of this procedure, Figure \ref{fig:so_pd} shows the photodissociation rate of 
SO as a function of radius for both internal and external photons for blackbody temperatures of 
2330\,K (blue) and 2600\,K 
(orange) and a mass-loss rate of 10$^{-7}$ M$_\odot$ yr$^{-1}$. Solid curves correspond to a 
smooth outflow, 
dashed curves to a one-component porosity model (effectively void inter-clump component), with 
$\fvol = 0.1$ and $l_* = 10^{13}$ cm, and dotted curves to a two-component model with $\fic = 0.5$, 
i.e. equal mass in the clump and inter-clump gas, and the same $\fvol$ and $l_*$ as the 
one-component model. 
The importance of internal photons to direct photodissociation of SO increases at the 
hotter temperature and dominates the interstellar rate  only inside 3--6 $\times$ 10$^{14}$ cm. 
Species with low bond energies can have rate coefficients due to internal photons substantially 
larger than their interstellar values whereas those with strong bonds, such as CO and N$_2$, are 
not affected by the presence of these internal photons.

\section{Results} \label{sec:results}

\begin{table*}
    \caption{Column density [cm$^{-2}]$ of CH$_4$, SO, NH$_3$ and H$_2$O in an C-rich outflow 
    for a smooth outflow 
    without and with internal photons and a one-component clumpy outflow ($\fvol = 0.1$ and 
    $l_* = 10^{13}$ cm) with internal photons. 
The corresponding abundance profiles are shown in Figure \ref{fig:ch4-so}.}
    \centering
    \begin{tabular}{c c c c c c c c }
    \hline \hline 
    	$\dot{M}$ & $T_*$ & \multicolumn{2}{c}{Smooth, no IP} & \multicolumn{2}{c}{Smooth, with IP} & \multicolumn{2}{c}{Clumpy, with IP} \\
    \cmidrule(lr){1-1} \cmidrule(lr){2-2} \cmidrule(lr){3-4} \cmidrule(lr){5-6}  \cmidrule(lr){7-8} 
 &  & CH$_4$ & SO & CH$_4$ & SO & CH$_4$ & SO \\ 
    \cmidrule(lr){1-1} \cmidrule(lr){2-2} \cmidrule(lr){3-4} \cmidrule(lr){5-6}  \cmidrule(lr){7-8} 
\raisebox{+0.4\normalbaselineskip}[0pt][0pt]{\multirow{3}{*}{\rotatebox[origin=c]{90}{\begin{tabular}[t]{c@{}c@{}}$10^{-5}$\\$\mathrm{M}_\odot\ \mathrm{yr}^{-1}$\end{tabular} }}} 
& 2100 K  & 1.56e+13 & 6.60e+11    & 1.56e+13 & 6.60e+11    & 2.02e+13 & 1.85e+12 \\
& 2330 K  & 2.32e+13 & 6.59e+11    & 2.32e+13 & 6.59e+11    & 1.34e+14 & 3.94e+12 \\
& 2600 K  & 3.34e+13 & 6.57e+11    & 3.34e+13 & 6.57e+11    & 1.43e+16 & 6.54e+13 \\
    \cmidrule(lr){1-1} \cmidrule(lr){2-2} \cmidrule(lr){3-4} \cmidrule(lr){5-6}  \cmidrule(lr){7-8} 
\raisebox{+0.5\normalbaselineskip}[0pt][0pt]{\multirow{3}{*}{\rotatebox[origin=lB]{90}{\begin{tabular}[t]{c@{}c@{}}$10^{-6}$\\$\mathrm{M}_\odot\ \mathrm{yr}^{-1}$\end{tabular} }}}
& 2100 K  & 1.28e+12 & 2.69e+11    & 1.28e+12 & 2.71e+11    & 5.46e+13 & 1.38e+12 \\
& 2330 K  & 2.02e+12 & 2.66e+11    & 2.48e+12 & 3.17e+11    & 1.15e+14 & 1.30e+12 \\
& 2600 K  & 3.01e+12 & 2.63e+11    & 1.44e+14 & 5.11e+11    & 1.67e+16 & 4.35e+11 \\
    \cmidrule(lr){1-1} \cmidrule(lr){2-2} \cmidrule(lr){3-4} \cmidrule(lr){5-6}  \cmidrule(lr){7-8} 
\raisebox{+0.5\normalbaselineskip}[0pt][0pt]{\multirow{3}{*}{\rotatebox[origin=c]{90}{\begin{tabular}[t]{c@{}c@{}}$10^{-7}$\\$\mathrm{M}_\odot\ \mathrm{yr}^{-1}$\end{tabular} }}}
& 2100 K  & 6.04e+12 & 4.28e+10    & 4.56e+12 & 7.94e+09    & 1.56e+14 & 1.67e+11 \\
& 2330 K  & 1.45e+13 & 4.33e+10    & 4.28e+12 & 3.63e+09    & 2.32e+14 & 6.38e+10 \\
& 2600 K  & 2.99e+13 & 4.40e+10    & 9.25e+11 & 9.02e+08    & 6.00e+14 & 9.67e+08 \\
    \cmidrule(lr){1-1} \cmidrule(lr){2-2} \cmidrule(lr){3-4} \cmidrule(lr){5-6}  \cmidrule(lr){7-8} 
 &  & NH$_3$ & H$_2$O & NH$_3$ & H$_2$O & NH$_3$ & H$_2$O \\ 
    \cmidrule(lr){1-1} \cmidrule(lr){2-2} \cmidrule(lr){3-4} \cmidrule(lr){5-6}  \cmidrule(lr){7-8} 
\raisebox{+0.4\normalbaselineskip}[0pt][0pt]{\multirow{3}{*}{\rotatebox[origin=c]{90}{\begin{tabular}[t]{c@{}c@{}}$10^{-5}$\\$\mathrm{M}_\odot\ \mathrm{yr}^{-1}$\end{tabular} }}} 
& 2100 K  & 5.06e+11 & 6.03e+13    & 5.06e+11 & 6.03e+13    & 6.57e+11 & 9.44e+13 \\
& 2330 K  & 5.65e+11 & 7.00e+13    & 5.65e+11 & 7.00e+13    & 7.56e+11 & 8.31e+14 \\
& 2600 K  & 6.27e+11 & 8.15e+13    & 6.27e+11 & 8.15e+13    & 8.48e+11 & 1.74e+14 \\
    \cmidrule(lr){1-1} \cmidrule(lr){2-2} \cmidrule(lr){3-4} \cmidrule(lr){5-6}  \cmidrule(lr){7-8} 
\raisebox{+0.5\normalbaselineskip}[0pt][0pt]{\multirow{3}{*}{\rotatebox[origin=lB]{90}{\begin{tabular}[t]{c@{}c@{}}$10^{-6}$\\$\mathrm{M}_\odot\ \mathrm{yr}^{-1}$\end{tabular} }}}
& 2100 K  & 2.29e+10 & 5.54e+12    & 2.29e+10 & 5.54e+12    & 2.24e+12 & 8.08e+14 \\
& 2330 K  & 2.67e+10 & 6.54e+12    & 2.67e+10 & 6.53e+12    & 4.10e+12 & 9.63e+14 \\
& 2600 K  & 3.25e+10 & 7.91e+12    & 3.35e+10 & 8.50e+12    & 1.80e+12 & 1.13e+15 \\
    \cmidrule(lr){1-1} \cmidrule(lr){2-2} \cmidrule(lr){3-4} \cmidrule(lr){5-6}  \cmidrule(lr){7-8} 
\raisebox{+0.5\normalbaselineskip}[0pt][0pt]{\multirow{3}{*}{\rotatebox[origin=c]{90}{\begin{tabular}[t]{c@{}c@{}}$10^{-7}$\\$\mathrm{M}_\odot\ \mathrm{yr}^{-1}$\end{tabular} }}}
& 2100 K  & 1.04e+10 & 2.43e+14    & 9.51e+09 & 2.42e+14    & 1.84e+12 & 1.56e+15 \\
& 2330 K  & 4.72e+10 & 3.36e+14    & 2.04e+10 & 3.29e+14    & 4.51e+12 & 1.79e+15 \\
& 2600 K  & 1.95e+11 & 4.59e+14    & 5.40e+09 & 3.21e+14    & 1.52e+12 & 1.46e+15 \\
    	\hline 	\hline
    \end{tabular}%
    \label{table:crich}
\end{table*}

We have made a set of calculations for a range of mass-loss rates, 10$^{-5}$--10$^{-7}$ 
M$_\odot$ yr$^{-1}$, and for both C-rich and O-rich AGB stars, with initial 
abundances given in \citet{agu10} and using the chemical model described by \citet{mce13}, 
publicly available through the UMIST Database for Astrochemistry website\footnote{\url{www.udfa.net}}, 
adapted by \citet{vds18a} to include a clumpy outflow. The calculations have been performed for a 
smooth outflow, a one-component outflow with $\fvol = 0.1$ and $l_* = 10^{13}$ cm, 
and a two-component with $\fic = 0.5$, $\fvol = 0.1$ and $l_* = 10^{13}$ cm.
We choose some representative species to discuss both the 
direct and indirect effects of internal UV on abundance distributions. 
The selection of these species is based on (i) their observability, (ii) their 
formations and destruction pathways are sensitive to the presence of internal photons, and 
(iii) some are `unexpected', e.g. H$_2$O in C-rich and HCN and CS in O-rich CSEs.
We concentrate here on the results at the two lower mass-loss rates and for the 
smooth and one-component clumpy outflows only. At the highest mass-loss rate, the 
effects of internal photons are less important because of the significantly higher 
extinction in this case (see \citeauthor{vds18a}~\citeyear{vds18a}). Similarly, the 
effects of internal photons in the two-component model are less pronounced due to its 
higher effective extinction than in the one-component model.

\subsection{Carbon-rich outflows}   \label{subsect:Crich}

Figure \ref{fig:ch4-so} shows the radial fractional abundances and Table~\ref{table:crich}
the radial column densities of CH$_4$, SO, NH$_3$ and H$_2$O for the C-rich case for 2100\,K 
(blue), 2330\,K (orange) and 2600\,K (green).
Solid curves refer to a smooth outflow without internal photons, 
dashed curves to a smooth outflow with internal photons, and dotted curves to a clumpy, one-component
outflow with internal photons.

The CH$_4$ abundance in the C-rich outflow can change by up to five orders of magnitude for 
$r$ less than 4 $\times$ 10$^{15}$ cm due to a combination of internal photons and clumpiness 
(Panel A, Figure \ref{fig:ch4-so}). 
Internal photons liberate carbon from the C-bearing parent species HCN and CS, with a larger 
increase in abundance for the higher stellar temperature.
Fast neutral-neutral reactions involving H$_2$ effectively hydrogenate carbon to form CH$_4$, 
transferring the increase in C abundance to CH$_4$.
The CH$_4$ radial behaviour changes for a clumpy outflow, since despite the larger C abundance 
due to photodissociation by interstellar photons of HCN and CS, the increased penetration of 
the internal photons now destroys the newly formed carbon hydrides.
The increased destruction of carbon hydrides by internal photons causes the decrease in CH$_4$ 
abundance when including internal photons at $\dot{M} = 10^{-7}$ M$_\odot$ yr$^{-1}$ 
(Panel B, Figure \ref{fig:ch4-so}). 
For a clumpy outflow, the larger C abundance in the inner regions compensates this effect, 
albeit only close to the star for the higher stellar temperature. At lower temperatures the 
dust extinction is sufficiently large for the carbon hydrides to form, leading to an overall 
larger CH$_4$ abundance.

For SO, the two main formation reactions are O + HS $\rightarrow$ SO + H and OH + S $\rightarrow$ SO + H. 
Which reaction dominates depends on the relative abundances of the reactants.
Photodissociation of the parent species SiS by internal photons liberates S close to the star. 
Hydrogenation of S leads to the formation of HS, increasing its abundance in the innermost 
region as well.
The SO radial behaviour hence roughly follows that of HS. 
For a higher stellar temperature and a clumpy or lower mass-loss rate outflow, the HS abundance 
decreases rapidly due 
to reactions with H. In such outflows, H is more abundantly present due to internal photons. 
In this case the OH + S reaction then takes over and the SO radial behaviour follows that of OH. 
OH is formed through the hydrogenation of O, which abundance increases slightly in the 
inner region due to photodissociation of the parent species SiO. CO is not affected by 
internal photons due to its small photodissociation rate.
In low mass-loss rate, high stellar temperature outflows, OH is formed in greater abundances close to the star 
thanks to the photodissociation of H$_2$O by internal and, for clumpy outflows, interstellar photons. 
The extent of its molecular shell is however limited due to the photodissociation of H$_2$O 
itself, which is reflected in the SO abundance profile.

For NH$_3$, internal photons affect the inner regions only for outflows with $\dot{M} = 10^{-6}$ 
M$_\odot$ yr$^{-1}$ but a more extended region for lower mass-loss rate outflows.
Smooth outflows with $\dot{M} = 10^{-6}$ M$_\odot$ yr$^{-1}$ have abundances which are essentially 
unaffected by the inclusion of internal photons but which, when both clumps and internal photons are 
present, increase by two to three orders of magnitude at $r$ smaller than 10$^{15}$ cm and are 
always larger than the smooth models at all radii (Panel E, Figure \ref{fig:ch4-so}).
In the clumpy outflow, NH$_3$ formation at $r \leq 10^{15}$ cm through the reactions 
H$_2$ + N $\rightarrow$ NH + H and O + HCN $\rightarrow$ NH + CO, followed by H atom 
abstraction reactions with H$_2$. Here O atoms are liberated through the photodissociation 
by internal photons of the parent species SiO. In the higher temperature models, the 
NH$_3$ abundance is lower due to the loss of O atoms through fast reaction with H$_2$. 
The behaviour of NH$_3$ in clumpy outflows and outflows with $\dot{M} = 10^{-7}$ 
M$_\odot$ yr$^{-1}$ is similar to that of CH$_4$, although its absolute abundance is 
generally lower due to its larger internal photodissociation rate.

H$_2$O is only affected by internal photons for lower mass-loss rate outflows (Panel H, 
Figure \ref{fig:ch4-so}). The increase seen in clumpy outflows is due to the clumpiness of 
the outflow, rather than the presence of internal photons.
The very small increase for $T_* = 2600$ K is caused by the larger O abundance close 
to the star.
For outflows with high stellar temperatures, OH is photodissociated close to the star by 
internal photons, leading a decrease in H$_2$O abundance closer to the star. 
This decline in abundance closer to the star also holds for the clumpy outflows with a 
high stellar temperature compared to the clumpy outflow without internal photons.

\subsection{Oxygen-rich outflows}       \label{subsect:orich}

\begin{figure*}[ht!]
\centering
\centering
\includegraphics[scale=0.4]{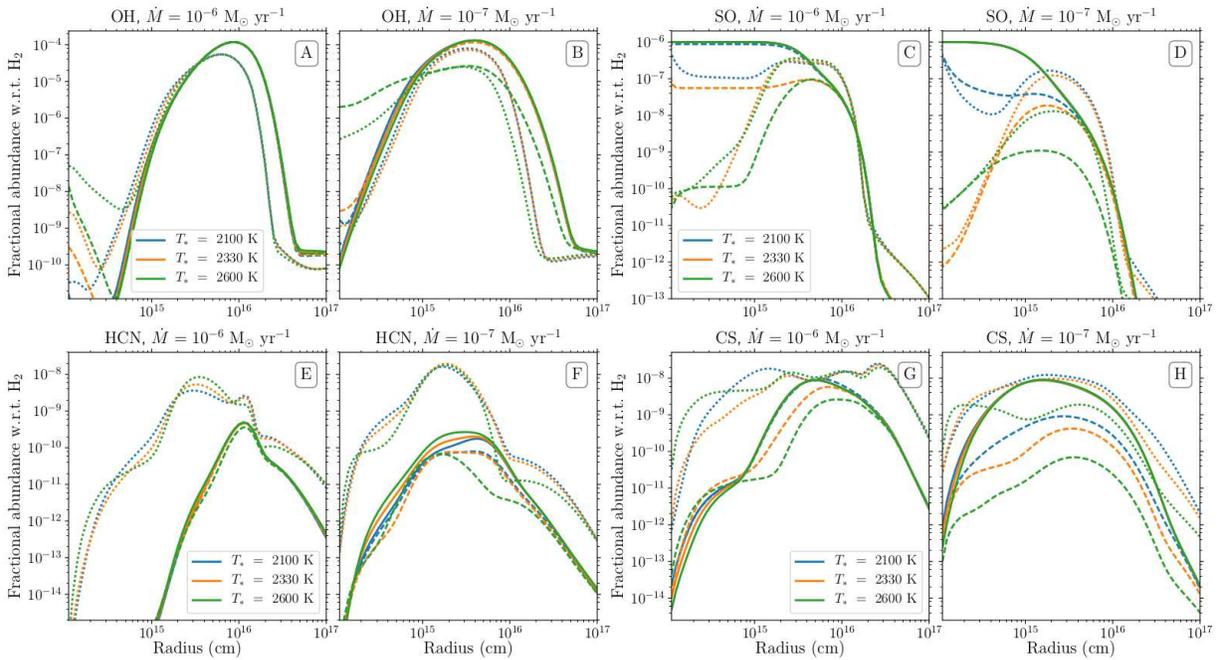}
\caption{Radial fractional abundance of OH, SO, HCN and CS for 2100\,K (blue), 
2330 K (orange) and 2600 K (green) in the O-rich case. Solid curves refer to a smooth outflow 
without internal photons, 
dashed curves to a smooth outflow with internal photons, and dotted curves to a clumpy, 
one-component outflow with internal photons. 
\label{fig:so-oh}}
\end{figure*}

Figure \ref{fig:so-oh} presents the abundances and Table~\ref{table:orich}
the radial column densities of OH, SO, HCN and NH$_3$ in the case of an O-rich outflow.
For the higher mass-loss rate of 10$^{-6}$ M$_\odot$ yr$^{-1}$, the production of OH is 
enhanced by internal photons inside 10$^{15}$ cm due to the photodissociation of parent 
H$_2$O, although we note that only around
1\% of water is destroyed in this region.  
At the lowest mass-loss rate, extinction is reduced and the higher UV flux at 
$T_*$ = 2600\,K results in the efficient photodissociation of OH so that 
its abundance is much reduced over the range 10$^{15}$--10$^{17}$ cm.
Despite its larger inner wind abundance, the radial profile of clumpy, 
lower mass-loss rate outflow follows that of H$_2$O as well: due to the large porosity of the 
outflow, the molecular envelope of H$_2$O only reaches to $\sim 10^{15}$ cm and gives rise to 
the large OH abundance inside 3-4 $\times$ 10$^{14}$ cm. 

\begin{table*}
    \caption{Column density [cm$^{-2}]$ of OH, SO, HCN and CS in an O-rich outflow for a smooth outflow without and with internal photons and a one-component clumpy outflow ($\fvol = 0.1$ and $l_* = 10^{13}$ cm) with internal photons. 
The corresponding abundance profiles are shown in Figure \ref{fig:so-oh}. 
    }
    \centering
    \begin{tabular}{c c c c c c c c }
    \hline \hline 
    \noalign{\smallskip}
    	$\dot{M}$ & $T_*$ & \multicolumn{2}{c}{Smooth, no IP} & \multicolumn{2}{c}{Smooth, with IP} & \multicolumn{2}{c}{Clumpy, with IP} \\
    \cmidrule(lr){1-1} \cmidrule(lr){2-2} \cmidrule(lr){3-4} \cmidrule(lr){5-6}  \cmidrule(lr){7-8} 
    	\noalign{\smallskip}
 &  & OH & SO & OH & SO & OH & SO \\ 
    \cmidrule(lr){1-1} \cmidrule(lr){2-2} \cmidrule(lr){3-4} \cmidrule(lr){5-6}  \cmidrule(lr){7-8} 
\raisebox{+0.4\normalbaselineskip}[0pt][0pt]{\multirow{3}{*}{\rotatebox[origin=c]{90}{\begin{tabular}[t]{c@{}c@{}}$10^{-5}$\\$\mathrm{M}_\odot\ \mathrm{yr}^{-1}$\end{tabular} }}} 
& 2100 K  & 2.54e+16 & 6.69e+16     & 2.54e+16 & 6.69e+16     & 2.33e+16 & 2.42e+16 \\
& 2330 K  & 2.55e+16 & 6.69e+16     & 2.55e+16 & 6.69e+16     & 2.35e+16 & 1.23e+14 \\
& 2600 K  & 2.53e+16 & 6.69e+16     & 2.53e+16 & 6.69e+16     & 2.29e+16 & 2.30e+14 \\
    	\noalign{\smallskip}
    \cmidrule(lr){1-1} \cmidrule(lr){2-2} \cmidrule(lr){3-4} \cmidrule(lr){5-6}  \cmidrule(lr){7-8} 
\raisebox{+0.5\normalbaselineskip}[0pt][0pt]{\multirow{3}{*}{\rotatebox[origin=lB]{90}{\begin{tabular}[t]{c@{}c@{}}$10^{-6}$\\$\mathrm{M}_\odot\ \mathrm{yr}^{-1}$\end{tabular} }}}
& 2100 K  & 1.62e+16 & 6.49e+15     & 1.62e+16 & 5.68e+15     & 1.14e+16 & 1.13e+15 \\
& 2330 K  & 1.62e+16 & 6.49e+15     & 1.62e+16 & 3.87e+14     & 1.09e+16 & 1.27e+14 \\
& 2600 K  & 1.61e+16 & 6.49e+15     & 1.61e+16 & 2.23e+13     & 1.04e+16 & 1.31e+14 \\
    \cmidrule(lr){1-1} \cmidrule(lr){2-2} \cmidrule(lr){3-4} \cmidrule(lr){5-6}  \cmidrule(lr){7-8} 
\raisebox{+0.5\normalbaselineskip}[0pt][0pt]{\multirow{3}{*}{\rotatebox[origin=c]{90}{\begin{tabular}[t]{c@{}c@{}}$10^{-7}$\\$\mathrm{M}_\odot\ \mathrm{yr}^{-1}$\end{tabular} }}}
& 2100 K  & 6.11e+15 & 5.71e+14     & 6.06e+15 & 7.30e+13     & 3.55e+15 & 5.38e+13 \\
& 2330 K  & 5.85e+15 & 5.71e+14     & 5.15e+15 & 1.22e+12     & 3.00e+15 & 6.31e+12 \\
& 2600 K  & 5.56e+15 & 5.71e+14     & 4.21e+15 & 1.87e+11     & 2.10e+15 & 9.57e+11 \\
    \cmidrule(lr){1-1} \cmidrule(lr){2-2} \cmidrule(lr){3-4} \cmidrule(lr){5-6}  \cmidrule(lr){7-8} 
    	\noalign{\smallskip}
 &  & HCN & CS & HCN & CS & HCN & CS \\ 
    \cmidrule(lr){1-1} \cmidrule(lr){2-2} \cmidrule(lr){3-4} \cmidrule(lr){5-6}  \cmidrule(lr){7-8} 
\raisebox{+0.4\normalbaselineskip}[0pt][0pt]{\multirow{3}{*}{\rotatebox[origin=c]{90}{\begin{tabular}[t]{c@{}c@{}}$10^{-5}$\\$\mathrm{M}_\odot\ \mathrm{yr}^{-1}$\end{tabular} }}} 
& 2100 K  & 4.97e+10 & 5.03e+12     & 4.97e+10 & 5.03e+12     & 5.45e+11 & 2.18e+13 \\
& 2330 K  & 5.01e+10 & 4.16e+12     & 5.01e+10 & 4.16e+12     & 5.45e+11 & 1.28e+13 \\
& 2600 K  & 5.31e+10 & 4.67e+12     & 5.31e+10 & 4.67e+12     & 5.29e+11 & 3.77e+13 \\
    	\noalign{\smallskip}
    \cmidrule(lr){1-1} \cmidrule(lr){2-2} \cmidrule(lr){3-4} \cmidrule(lr){5-6}  \cmidrule(lr){7-8} 
\raisebox{+0.5\normalbaselineskip}[0pt][0pt]{\multirow{3}{*}{\rotatebox[origin=lB]{90}{\begin{tabular}[t]{c@{}c@{}}$10^{-6}$\\$\mathrm{M}_\odot\ \mathrm{yr}^{-1}$\end{tabular} }}}
& 2100 K  & 2.48e+10 & 2.14e+12     & 2.49e+10 & 2.21e+12     & 1.35e+12 & 2.06e+13 \\
& 2330 K  & 2.52e+10 & 2.14e+12     & 2.37e+10 & 1.03e+12     & 1.55e+12 & 8.86e+12 \\
& 2600 K  & 2.57e+10 & 2.14e+12     & 1.76e+10 & 4.16e+11     & 2.01e+12 & 2.02e+13 \\
    \cmidrule(lr){1-1} \cmidrule(lr){2-2} \cmidrule(lr){3-4} \cmidrule(lr){5-6}  \cmidrule(lr){7-8} 
\raisebox{+0.5\normalbaselineskip}[0pt][0pt]{\multirow{3}{*}{\rotatebox[origin=c]{90}{\begin{tabular}[t]{c@{}c@{}}$10^{-7}$\\$\mathrm{M}_\odot\ \mathrm{yr}^{-1}$\end{tabular} }}}
& 2100 K  & 7.77e+09 & 1.08e+12     & 5.04e+09 & 9.18e+10     & 9.84e+11 & 1.95e+12 \\
& 2330 K  & 1.04e+10 & 1.07e+12     & 4.70e+09 & 3.02e+10     & 1.01e+12 & 1.92e+12 \\
& 2600 K  & 1.50e+10 & 1.05e+12     & 4.38e+09 & 4.63e+09     & 8.82e+11 & 8.86e+11 \\
\hline 	\hline
    \end{tabular}%
    \label{table:orich}
\end{table*}

SO is included as a parent species in the O-rich outflow. Internal photons photodissociate 
SO in the innermost region, leading to a decrease in its abundance of up to five orders of 
magnitude (Panel D, Figure \ref{fig:so-oh}). Its destruction is larger for a higher stellar temperature.
In case of a clumpy or lower $\dot{M}$ outflow, the SO abundance close to the star is however 
larger for higher $T_*$. This is due to the larger OH abundance in this region for these outflows, 
which replenishes the SO abundance through the reaction OH + S $\rightarrow$ SO + H. 

The large increase in HCN abundance seen for clumpy outflows, (Panel E, Figure \ref{fig:so-oh}), 
is mainly due to the clumpiness of the outflow rather than due to internal photons. 
For $r$ smaller than $10^{15}$ cm, the decrease in abundance  is due to photodissociation by 
internal photons. 
Around $10^{15}$ cm, its main production reaction shifts from H$_2$ + CN $\rightarrow$ HCN + H 
to N + CH$_2$ $\rightarrow$ HCN + H. 
The CH$_2$ abundance is smaller for higher stellar temperatures, as internal photons more 
readily destroy hydrogenated carbon, leading to a decline in the peak HCN abundance, which 
is larger for lower mass-loss rate outflows. 

As for HCN, the large increase in the CS abundance for the clumpy outflows is again mainly 
the result of clumpiness rather than internal photons. 
For the lower mass-loss rate outflows, the peak CS abundance decreases up to two orders of 
magnitude when including internal photons in a clumpy or smooth outflow.
At radial distances less than $10^{15}$ cm, the main reaction forming CS is 
HCS$^+$ + e$^-$ $\rightarrow$ CS + H, where HCS$^+$ is mostly formed via 
CH$_3^+$ + S $\rightarrow$ HCS$^+$ + H$_2$. Including internal photons leads to a 
larger S abundance close to the star, due to the photodissociation of the parent 
species H$_2$S. This then leads to a larger HCS$^+$ abundance, and therefore CS abundance.
In clumpy outflows with 10$^{-7}$ M$_\odot$ yr$^{-1}$, the CS abundance decreases 
before $10^{15}$ cm (Panel F, Figure \ref{fig:so-oh}). This is linked to the increased 
O abundance close to the star when including internal photons, destroying CS via the 
reaction O + CS $\rightarrow$ S + CO. Beyond $10^{15}$ cm, the main reaction forming CS 
shifts to C + SO $\rightarrow$ CS + O. 
The influence of internal photons on the SO abundance is hence propagated to the 
CS abundance profile, causing a decrease in peak abundance of up to two orders of magnitude.

\section{Discussion} \label{discussion}

Our calculations show that the internal UV photons in combination with a clumpy outflow can 
have a significant effect on the chemistry and radial distributions of specific molecules 
in both C-rich and O-rich AGB envelopes, in particular, forming the `unexpected' molecules 
seen in these envelopes (see Section \ref{sec:intro}).

For the largest mass-loss rate considered, 10$^{-5}$ M$_\odot$ yr$^{-1}$, we find that the 
dust extinction renders these photons unimportant for the blackbody temperatures we have 
assumed here.  
Inner photons hence cannot account for the observed column density of NH$_3$ of the high mass-loss 
rate C-rich AGB star IRC+10216 \citep{sch16}, one its `unexpected' species.
For lower mass-loss rates, we find our clumpy models can give rise to significantly enhanced 
abundances of several species in the inner CSE, around 10$^{15}$ cm.  For example, in our 
C-rich calculations, we find water abundances on the order of (3-60) $\times$ 10$^{-7}$, 
consistent with the abundances found by \citet{lom16} in their study of C-rich AGB stars, 
while the H$_2$O column density is enhanced by one to two orders of magnitude for effective 
temperatures of 2100\,K and 2330\,K (Table \ref{table:crich}). 
For O-rich outflows, we find that HCN reaches peak abundances of around 10$^{-8}$, less 
than the peak abundances of around 5 $\times$ 10$^{-7}$ observed in IK Tau \citep{dec10a} and 
R Dor \citep{vds18c}. Although these peak abundances fall below those observed, we find 
that the HCN column densities are much enhanced, by around one to two orders of magnitude 
when internal photons are included, (Table \ref{table:orich}), even for mass-loss rates as 
large as 10$^{-5}$ M$_\odot$ yr$^{-1}$.

Our models do have limitations. In addition to the uncertainty associated with calculating 
the photodissociation rates of species for which no wavelength-dependent cross-sections are 
available, much of our gas-phase chemistry in the inner envelope occurs at high temperatures 
where two-body reactions are less well studied and hence may raise issues of completeness of 
chemistry. 
Although not shown here, two-component clumpy models generally show smaller effects of 
UV photons than the one-component model presented here.
Such models have higher effective extinction since the inter-clump medium is not void and 
would diminish the effects of internal photons.
Additionally, as noted in Section \ref{sec:intro}, many AGB stars may contain binary 
companions. The presence of harder radiation fields from a hot binary or associated 
accretion disk might also be expected to change results appreciably.
Finally, we have used a blackbody spectrum to calculate the UV photons emitted by the AGB 
star due to the limited ranges of stellar temperatures and surface gravities covered by 
stellar atmosphere models. Blackbody spectra however do not take into account the effects 
of molecular bands on the emitted UV spectrum and introduce a potentially significant 
uncertainty into our results.

\section{Conclusions} \label{conclusions}

We have presented the first model calculations that include the chemical effects of internal 
UV photons, here assumed to be blackbody radiation from cool AGB stars using the 
porosity formalism introduced by \citet{vds18a} and the CSE chemistry code from \citet{mce13}. 
For the radiation fields used here, we find that internal photons are essentially 
unimportant at 10$^{-5}$ M$_\odot$ yr$^{-1}$ due to the large value of dust extinction.

Our results show that at lower mass-loss rates, for a clumpy circumstellar medium close to 
the photosphere, as is observed for several AGB stars, these UV photons are capable of 
influencing the chemistry that occurs deep in the CSE. These effects are more significant the 
larger the effective temperature and, in general, the lower the mass-loss rate. 
The distribution of molecules can be altered either directly through enhanced photodissociation 
rates or indirectly through changes to the abundances of other species involved in their 
formation or destruction. 
We find that the radial distribution of the fractional abundance of H$_2$O in C-rich outflows 
is significantly enhanced in the inner envelope while its column density and that of HCN in 
O-rich outflows can increase by 1 to 2 orders of magnitude when internal photons and clumps 
are included in models. 

In future work, we shall consider the influence of harder radiation fields but it is clear 
that chemical kinetic models of the inner winds of AGB stars, including those of dust 
formation and shock chemistry following stellar pulsations, should include the photochemistry 
driven by these internal photons.

\acknowledgments

We thank the anonymous referee for providing useful comments which helped improve this article.
TJM is grateful to the STFC for support through grant ST/P000321/1 and for the hospitality 
of Leiden Observatory where much of this work was carried out. MVdS acknowledges support from 
the Research Foundation -- Flanders through grant 12X6419N.

\bibliography{ip_arxiv}



\end{document}